\documentstyle{l-aa}
\input epsf.sty
\begin{document}
\thesaurus{03 - 11.01.2; 11.14.1; 02.18.6; 13.21.1; 13-25-2}
\title{Cloud model of the mean quasar spectrum}
     
\author{ Bo\. zena Czerny \inst{1},  
Anne-Marie Dumont \inst{2}}
\offprints{A.-M. Dumont}
     
\institute{{$^1$}Copernicus Astronomical Center, Bartycka 18. PL-00-716 Warsaw, 
Poland \\ {$^2$}DAEC, Observatoire de Paris, Section de Meudon, 92195 Meudon, 
France \\} 

\date{Received 15 December 1997; accepted 15 June 1998} 

\maketitle

 \markboth{B. Czerny, A.-M. Dumont; Cloud model of the mean quasar spectrum}{}

\begin{abstract}

We assume a distribution of clouds optically thick for electron scattering
(OTCM) which are moderately optically thin for absorption and we consider 
them as a model of the mean quasar spectrum of Laor et al. (1997). 
We show that the model is particularly
sensitive to the value of the ionization parameter $\xi$ and that for 
$\xi \sim 500$ the model well reproduces the optical/UV/X-ray mean quasar
spectrum, in agreement with the estimates of the ionization parameter based 
on the energy of the iron $K_{\alpha}$ line. We cannot definitively reject
synchrotron emission as a source of primary radiation but we favor the model
in which the hard X-ray emission is produced by Compton scattering of soft 
photons in a central hot medium surrounded by cool clouds. In such a model
clouds are located typically at the distance of $\sim 12 R_{Schw}$, with the
covering factor about 0.88 and the radius of hot plasma is $\sim 9 R_{Schw}$.
The model explains optical/UV emission as predominantly due to the dark
sides of the clouds and the soft X-ray emission as due to the reflection
by the irradiated sides of the clouds. Therefore, atomic features are expected
in these bands although they are hardly present in the observational data.
The kinematical effects connected with the cloud motion
affect those features but do not remove them. The level of the primary
emission required to model the mean quasar spectrum is too low to 
reproduce the equivalent width of the iron K$_{\alpha}$ line correctly
but more detailed computations may resolve this problem.

\end{abstract}


     
\section{Introduction} 

It is broadly accepted now that active galactic nuclei are powered by
accretion onto a supermassive black hole and that the flow is a multi-phase
(at least two-phase) medium. A hot optically thin phase is responsible for the 
X-ray and $\gamma$-ray emission while a cool phase of considerable optical 
depth is responsible for the thermal optical/UV emission as well as for the
preprocessing of a significant fraction of the radiation produced by the hot
phase (for a review, see e.g. Mushotzky et al. 1993). There is 
mounting evidence based on the shape of $K{\alpha}$ line profile as well as
on the variability in soft and hard X-rays, that the release of the 
major part of the X-ray
energy, of the thermal EUV/soft X-ray emission and the reprocessing of X-ray
by cold material takes place in a very compact region, of order of $\sim 10$
Schwarzschild radii and the cold matter is approximately in Keplerian
motion (see e.g. Nandra et al. 1997a). 
 
However, the geometrical arrangement of the two phases is still under debate.
The three most 
viable models are: a disk with a hot corona, very optically thick
clouds embedded in a hot medium, and moderately optically thick clouds also
coexisting with a hot medium. In order to differentiate between these models
detailed studies of all of them are necessary. The disk model 
is the best studied so
far (e.g. Ross \& Fabian 1993, Stern et al. 1995, Sincell \& Krolik 1997). 
Optically thick clouds were studied for example by Sivron \& Tsuruta 
(1993; see also the references therein). Clouds of 
moderate optical depth were broadly advertised as a 'free-free' 
emission model (e.g. Antonucci \& Barvainis 1988). 
However, clouds which are optically thin for electron
scattering are not a viable model for an AGN (Barvainis 1993, 
Collin-Souffrin et al. 1996, Kuncic et al. 1997; see
also Lightman \& White 1988). On the other hand clouds, which are 
optically thick for
electron scattering but optically thin for absorption in the UV and moderately
optically thick in the optical band 
(hereafter OTCM model), can reproduce the required fraction of
the X-ray emission and roughly produce the correct multi-wavelength
spectra, as well as  provide enough
material to support the nuclear activity (Collin-Souffrin et al. 1996;
hereafter Paper I). 
As the model may have an advantage over 
clouds optically thick for absorption it deserves a detailed study of its 
observational consequences.

In this paper we concentrate on the comparison of the predictions of
the OTCM with the broad band mean
spectrum for radio quiet quasars determined by Laor et al. (1997). 
As direct analysis of the X-ray emission does not
give any reliable answer to the question of the 
nature of the X-ray and $\gamma$-ray emission  due to the still poor 
determination of
the spectra of radio quiet AGN in the  $\sim 1 $MeV range 
(see e.g. Gondek et al. 1997 for combined Seyfert 1 spectra from GRO) 
we analyze three representative cases of the sources of emission external
to the clouds. As is
customary and convenient, we call this incident radiation for the clouds the
``primary'' radiation although this is unjustified when 
the X-ray emission is itself the emission of the clouds comptonized by a hot
medium.

\section{The model}

The overall outline of the cloud model and of the radiation transfer within a
cloud was presented in Paper I so we only 
summarize here the basic assumptions.

We consider a distribution of clouds surrounding a hot plasma which is a source
of incident (primary) radiation for the clouds. Sides of the clouds exposed to the
central source are bright while the unexposed sides are relatively dark and cold
because of the considerable optical depth of the clouds. The schematic geometry
is shown in Figs. 1 and 2.

All clouds are assumed to 
have the same density and to be exposed to the same incident flux since any
assumption about the radial distribution of the cloud properties would be
completely arbitrary without an underlying dynamical model. However, we
carefully calculate the radiative transfer in the clouds. The resulting broad
band spectra from the IR to X-rays are combinations of the radiation 
transmitted, emitted and scattered by clouds as well as a fraction of the 
primary radiation. The relative weight of these components depends on the
covering factor and the size of the source of primary emission. 

The radiation transfer code used in this paper (Dumont \& Collin-Souffrin, 
in preparation) has been improved in comparison with the one used in Paper I.
The ionization of all hydrogenic ions 
(not only hydrogen 
itself) proceeds from all 5 levels and interlocking (subordinate) lines are
included. 
These modifications influence to some extent the line emission and 
the Lyman edge, particularly due to its coincidence with 
the second level of He$^+$. 

Since the radiative transfer code of Dumont \& Collin-Souffrin is constrained
in the present version to photons with energy below 24 keV (although
Compton heating by harder X-rays is included) we calculate the shape of
the reflected 
spectrum above 24 keV using the method given by Lightman \& White (1988). We
use the opacities for the neutral gas since the ionization level is 
unimportant for these high energy photons. The continuity between the two
reflected spectra was achieved by adjusting the opacity parameter C of
Lightman \& White (1988) in their Eq. (17). The value of this
parameter was in all cases about $0.5 \times 10^{-5}$, by a factor 
of two lower than the value
suggested by George (private communication) for a neutral medium and elemental
abundances used in George \& Fabian (1991).

\begin{figure}
\epsfxsize = 80 mm \epsfbox[0 0 400 450]  {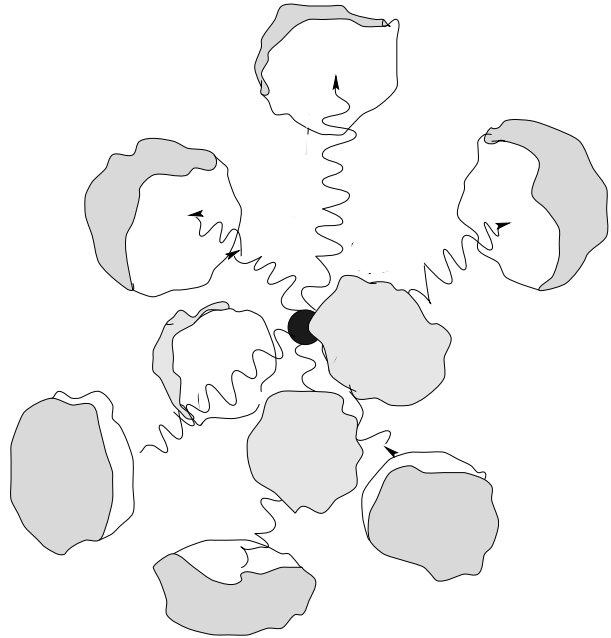}
\caption
{The overall view of the geometry of the model (A) and (B). Clouds surround
a central source of unspecified origin, with one of the clouds just 
accidentally partially blocking our direct line of sight to the source of 
primary emission.} 
\end{figure}

\begin{figure}
\epsfxsize = 80 mm \epsfbox[0 0 400 450]  {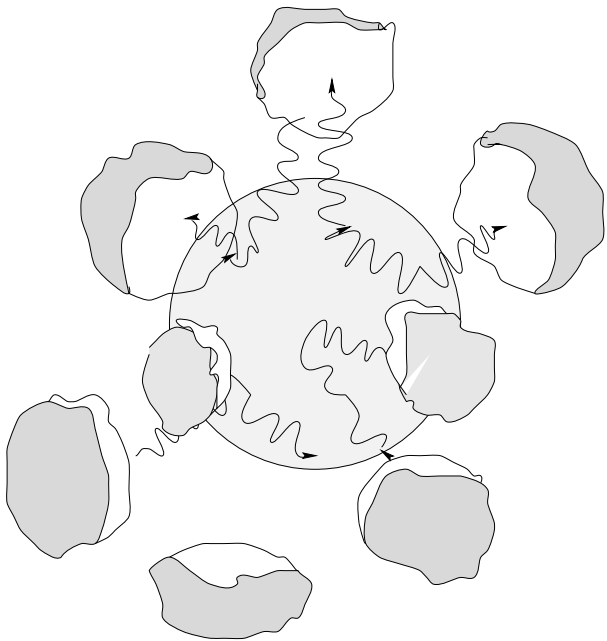}
\caption
{The overall view of the geometry of the 
the model (C). Hot plasma in the center acts as a source of 'primary 
emission' by Compton scattering soft photons emitted or reflected by the 
clouds.} 
\end{figure}

\subsection{Cloud parameters} 

Since, at this stage, we do not introduce a full dynamical model describing 
the cloud formation and disruption and their motion the radial distribution
of the clouds themselves and of their properties is arbitrary. Therefore, in
order to keep the parameterization as simple as possible we assume that all
clouds have the same (constant throughout a cloud) density $n$ and
column density $N_H$. Since clouds optically thin for electron scattering were
ruled out as a model for optical/UV/soft X-ray continuum we study clouds with
$log(N_H)\ge 25$, usually concentrating on the $log(N_H)= 26$ case. Higher
$N_H$ strongly suppress the emission from the unilluminated side of the cloud
so only reflection is seen. 
The clouds are opaque for the X-ray radiation
so even if some clouds are in our line of sight towards the primary X-ray 
source this would not lead to the presence of the transmitted, 
strongly absorbed spectral component. The primary emission intercepted by 
the clouds is partially reflected and partially reemitted by the dark 
side of the 
cloud (i.e. the side opposite to the central source) in the form of 
thermal emission.

The distribution of the clouds is described by the
covering factor $\Omega/4 \pi$ with respect to the X-ray source. If 
$\Omega/4 \pi$ is close to 1, the distribution necessarily is almost 
spherical. If $\Omega/4 \pi$ is smaller (e.g. 0.5) then the distribution may
either be still spherical, or rather constrained to the symmetry plane. These
two cases are different if Doppler effects connected with the
motion of the clouds are taken into account since plane confinement would 
result in broader features for the same velocity field.

\subsection{The primary radiation} 

The nature of the primary emission is still not known. In the past, non-thermal
models were favored. In such models a (usually power law) distribution of
relativistic electrons produces the radiation either by direct synchrotron
emission or by Comptonization of soft photons. These soft photons were 
either their
own synchrotron photons (SSC) or came from external source like an accretion
disk.
In these models the pair creation process was frequently important, 
particularly
for the shape of the spectrum above $\sim 1 MeV$. 

These models are still very popular in the case of radio loud AGN. However,
in the studies of radio quiet AGN the attention recently
shifted towards thermal models. In these models the hot plasma is
thermal, at a temperature of order of a few hundreds of keV, and it produces
the radiation through Comptonization of soft external photons. In these models
the effect of pair creation is usually negligible. 

These two basic families of models differ also in the extension of the primary
component into the low energy range. In non-thermal models the primary emission
may (but does not necessarily have to) extend down to the far-IR/mm band. It
was an attractive possibility in the late 70s and 80s 
when IR emission was generally
thought to be dominated by non-thermal processes. Evidence of a significant 
contribution of dust to the IR emission as well as some arguments based on
variability (e.g. Done et al. 1990 for NGC 4051) diminished the popularity of
that view. However, a single underlying
IR/X-ray continuum is still sometimes advocated on the basis of observational
arguments (e.g. Walter \& Fink 1993, Loska \& Czerny 1990, Fiore et al. 1995). 
Thermal models, on the other hand,
predict a contribution of the hard component only above the energy of the
seed photons for Comptonization, i.e. mostly starting from the UV band.

As a result, in their computations of the overall continua, various authors 
used different
specific assumptions about the shapes of the incident spectra for 
reprocessing. 
Lightman \& White (1988) consider a non-thermal model with an energy index
0.7 and a spectrum extending from 1 eV to 3 MeV, Guilbert \& Rees (1988)
have the primary non-thermal spectrum with energy index 0.5 (they also 
include the effect of pair creation on the shape of the primary). 
Sivron \& Tsuruta (1993) assume that the initial synchrotron radiation with
an energy index 1 extending from 0.1 eV  to 1 MeV is filtered by CFR cloudlets
(Celotti et al. 1992). This special, very compact, population of 
cloudlets, with densities $\sim 10^{16-18}$cm$^{-3}$, hydrogen column  $N_H \sim 
10^{21}$cm$^{2}$
and covering factor 1, absorbs the IR emission (through the free-free 
process) but it is transparent at the optical band and above. The
absorbed energy is reemitted in the form of free-free emission with the
adopted value of the temperature $10^6$ K. Such a preprocessed spectrum serves
as incident spectrum for the main clouds occupying the space between $R_{in}
= 10 R_{Schw}$ and $R_{out} = 1000 R_{Schw}$.

In order to study the influence of the shape of the incident flux on
the resulting spectrum predicted by the model we choose three representative
shapes of the primary continuum justified by general theoretical arguments. 
They are named models (A), (B) and (C), and they are shown in Fig. 3.

\subsubsection{Model A: synchrotron emission extending to IR} 

This incident radiation resembles closely the incident radiation 
adopted in our initial
study (Paper I) as well as by Kuncic et al. (1997). We assumed a low frequency
cut-off at 0.1eV as in Paper I. In order to reproduce 
well the observed 
spectrum (according the present knowledge) we assume an energy index either 
0.9,
after the classical paper on Seyfert 1 galaxies 
(Pounds et al. 1990) or steeper since in the case of quasars the 
situation is not clear: Laor et al. (1997) give a slope of 1 for radio quiet
objects while Williams et al. (1992) give 0.92 when flat spectrum objects are
excluded. We assume the values of the cutoff energy 100 keV and 280 keV 
since this last value
seems to be suggested by GRO data for Seyfert 1 galaxies (Grandi et al. 
1998 for MCG8-11-11; Madejski et al. 1995 for IC 44329A, Gondek et al. 1997
for a composite spectrum). No direct data constraints for cut-off energy 
in radio quiet quasars are available.

Such a model corresponds to the presence of a relatively strong magnetic
field within the hot plasma, such that the energy density of the magnetic field
is larger than the energy density of the soft photons available due to the
presence of cool clouds. Models of that type were studied e.g. by 
Maraschi et al. (1982), with lower frequency cut-off $10^{12} - 10^{13}$
Hz determined by self--absorption. 

The hypothesis 
of the existence of such an IR/X-ray power law was observationally tested in the 
case of the source NGC 4051 (Done et al. 1990) and its existence 
was not confirmed since
strong X-ray variability in this source was not accompanied by any optical
variability. On the other hand NGC 4051 is not a typical example of
an AGN and its optical emission may be strongly dominated by starlight. 
A number of other sources like NGC 5548 
(Korista et al. 1995) and
NGC 4151 (Edelson et al. 1997) 
show coherent day to day variations in optical, UV and X-ray band, 
with unmeasurable delays smaller than several hours. 

We do not analyze the conditions for
the production of such a primary emission in the present paper. We
simply assume the parametrization by a single power law and we fix the
low energy cut-off at 0.1 eV, as in Paper I.   

The input model parameters are: the ionization parameter, $\xi=L/nR^2$
(where $L$ is the bolometric luminosity), the cloud 
density, $n$, and
its hydrogen density column, $N_H$, (or cloud size). 
The observational appearance of the system is further parameterized by the
relative contribution of the primary (or incident) radiation  and 
the emission from dark sides of clouds with respect to the reflected
component. For purely random cloud distribution both parameters are uniquely
determined by the  covering factor, $\Omega /4 \pi$ (see eq. 2).

\subsubsection{Model B: synchrotron emission extending to UV} 

Our second model is a power law characterized by the same energy index and high energy
cut-off as before but the adopted low frequency cut-off is set at 30 eV
i.e. in the UV band ( $7 \times 10^{15}$ Hz). This simple model is
supposed to represent a situation when the gas is not simply in equipartition 
with the magnetic field but dominates the behaviour of the matter 
(Rees 1987). As the expected magnetic field is in this case two orders of
magnitude or more higher than in the previous case ($ B \sim 10^4 - 10^5$ 
gauss, see Celotti et al. 1992) the self-absorption frequency is 
proportionally higher. Additionally, in such a strong magnetic field a new
family of cloudlets, or filaments, may form in the innermost part of the flow.
Such cloudlets are opaque to photons
below, again, $10^{15}$ Hz due to their high density and free-free absorption
so they filter the primary emission before this emission may reach the main
clouds outside (Sivron \& Tsuruta 1993). Therefore the incident radiation
in the case of magnetic field dominating innermost flow  would not extend 
beyond
optical/UV. Absorption by cloudlets leads additionally to reemission of
the absorbed radiation in the form of a black body radiation. However, as
we cannot predict the temperature of such radiation and the covering factor
by the inner cloudlets we simply neglect this component in the present
consideration.

The input model parameters are the same as in the previous section.

\subsubsection{Model C: self-consistent thermal model} 

This model is based on the assumption that amplification of the magnetic field
within the flow is not efficient. Therefore, clouds are the only source of the
soft photons. We also assume this time that the hot medium is thermal, i.e.
basically characterized by the optical depth and the temperature. Hard X-ray
emission in this model forms by Compton upscatter of a fraction of the
soft photons from the
clouds while soft photons result from interception of a fraction of hard 
X-ray emission by clouds. 

In order to obtain  the appropriate shape of the hard X-ray continuum as 
described above we have to adopt a temperature of the hot medium equal
to  $\sim
3 \times 10^9$ K and an optical depth $\sim 0.1$. Adopting higher
(lower) value for the temperature would require a lower (higher) value of the
optical depth of the hot medium. The Compton parameter $y$ for such a plasma
is 1.1.

The fraction of
the soft photons intercepted by the hot medium results from the model
(see Paper I) and is approximately given by 
\begin{equation} 
({R_X \over R_{UV}})^2 = { 4 \pi \over \Omega f_X exp(y)}
\end{equation}
where $f_X$ is the 
fraction of photons reflected or reemitted by the illuminated
side of a cloud, determined by the physical conditions in the gas. 
Our model is therefore more flexible than the original
version of the corona model (Haardt \& Maraschi 1991) which requires 
$\Omega/4\pi$ equal to 0.5 and $R_X = R_{UV}$, thus practically fixing the value
of the Compton parameter $y$.  Recently developed clumpy corona models are
based on an additional covering factor for the hot medium which is equivalent
to our $R_X/R_{UV}$ ratio from the point of view of the efficiency of the 
Comptonization but the two models nevertheless differ with respect to
overall geometry and its (angular-dependent) appearance. 

The input parameters of the model are, as usual, the ionization parameter, the cloud 
density,
its hydrogen density column (or size), and the weights of the primary and
dark side components.

\begin{figure}
\epsfxsize = 80 mm \epsfbox[50 180 530 660] {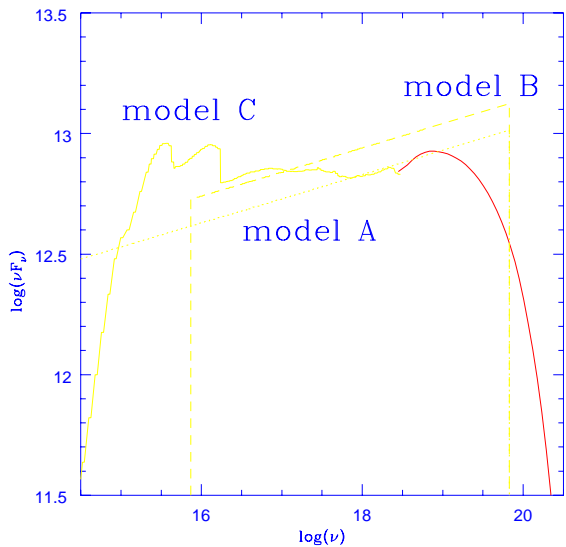}
\caption
{Three examples of the IR-Xray incident (primary) spectra for OTCM 
models, as described  in Section 2.2: model (A) - dotted line, model (B) - 
dashed line and model (C) - continuous line. Model C was computed assuming
the hot medium temperature equal to  $3 \times 10^9$ K and its optical depth
equal to  0.1.} 
\end{figure}

However, the computations of the model require an iterative procedure to 
achieve a
self-consistent solution for the soft and hard emission. In order to avoid
an arbitrarily adopted spectral shape in the optical band we start with
a mechanically heated cloud with temperature $\sim 10^5$ K. 
Next we compute the 
Comptonized spectrum which results from the interception of soft photons by hot 
plasma with the parameters given above. For that purpose we use the method of
Czerny \& Zbyszewska (1991), appropriate for a thin plasma. 
This radiation now serves as incident
radiation. Next we calculate the {\bf reflected} (not emitted) component as
the cloud faces the hot medium clearly with its hot hemisphere. The 
comptonization of this reflected component gives a second approximation to the 
hard X-ray emission spectrum and the model is iterated until it converges.
The normalization is fixed by the adopted value of the ionization parameter
and the ratio of the Comptonized reflected radiation to renormalized radiation
gives us accurately the fraction of soft photons intercepted by the hot gas, or
the relative extension of the two regions, roughly estimated by Eq. (1).

\section {Results} 

The observed spectrum of an AGN within the frame of the OTCM 
is in general a sum of the primary, a
reflected component and an emission from the dark sides  of the clouds, 
with the relative weight of the last two given by the
covering factor (see  Paper I). We make first a general 
qualitative discussion without
adopting any specific value of this parameter (Sect. 3.1). More detailed
discussion, however, clearly involves the covering
factor as additional parameter of the model (Sect. 3.2 and 3.3).

\subsection{General properties of the primary, reflected and emitted 
components}

\subsubsection{The choice of the cloud parameters}

\begin{figure}
\epsfxsize = 80 mm \epsfbox[50 180 530 660] {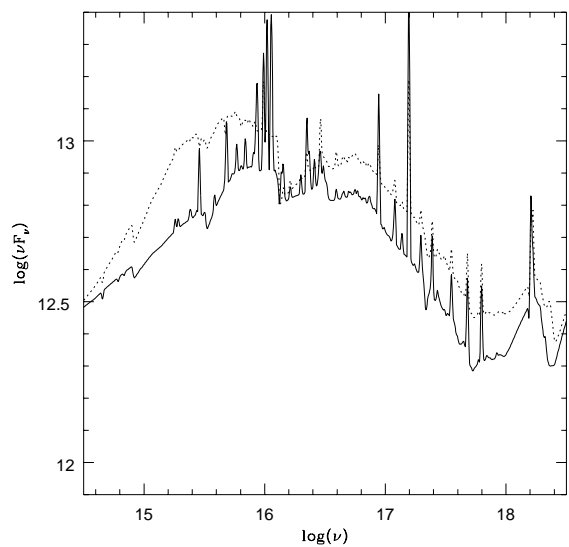}
\caption
{Two examples of the optical/X-ray spectrum coming from the bright sides of 
clouds calculated for model (A) assuming the cloud column density
 $10^{25}$ cm$^{-2}$ (continuous line) and $10^{26}$ cm$^{-2}$ (dotted line).  
Other parameters: ionization parameter $\xi = 1000$, extension of the incident
spectrum from 0.1 eV to 280 keV, the slope of the incident spectrum 
 $\alpha = 0.9 $ and  cloud density $10^{12}$
cm$^{-3}$.
} 
\end{figure}

\begin{figure}
\epsfxsize = 80 mm \epsfbox[50 180 530 660] {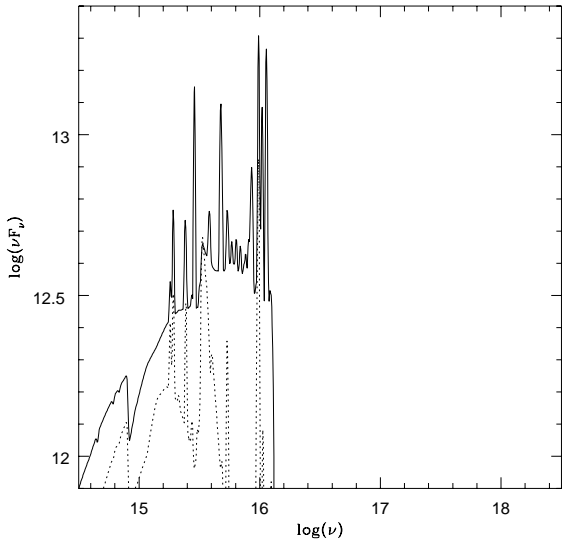}
\caption
{Two examples of the optical/X-ray spectrum coming from the dark sides of 
clouds calculated for model (A) assuming the cloud column density
 $10^{25}$ cm$^{-2}$ (continuous line) and $10^{26}$ cm$^{-2}$ (dotted line).  
Other parameters: ionization parameter $\xi = 1000$, extension of the incident
spectrum from 0.1 eV to 280 keV, the slope of the incident spectrum 
 $\alpha = 0.9 $ and  cloud density $10^{12}$
cm$^{-3}$. } 
\end{figure}

The clouds are parameterized by the density, $n$, and the hydrogen density
column, $N_H$. The reflected radiation, i.e. the radiation of the bright
illuminated sides of the clouds is not very sensitive to the column density as
long as the total optical depth of a cloud is high and in this paper we
deal with such a clouds. In Fig. 4 we show two examples of the 
reflected spectra for
two values of the column density (for model A). We see that the spectrum
is softer in the optical band and slightly harder in the soft X-rays. 
The column density
is important, however, for determination of the emission of the dark
sides of the clouds (see Fig. 5).  Larger column density leads to almost
a black body type of emission although atomic features are clearly visible
while lower column density is characterized by very strong
atomic features superimposed on the continuum.  The comparison with optically
thin clouds was already discussed in Paper I.

Examples of the spectra calculated for three values of the ionization parameter
$\xi$ (300, 1000 and 3000) were shown in Paper I.

The dependence on the density of the clouds is weak since we parameterize
the incident radiation not by flux but by the ionization parameter $\xi$.
Clouds with lower density ($10^{10}$ cm$^{-3}$) are characterized by slightly
bigger hydrogen and helium edges. Clouds with larger densities ($10^{14}$ 
cm$^{-3}$) are slightly optically thicker in the optical range thus leading
to lower emission in those wavelengths.

\subsubsection{The high energy extension and the slope of the primary emission}

We tested two values of the high energy extension of the incident spectrum:
100 keV and 280 keV. The effective change of the bolometric luminosity required
to preserve the value of the ionization parameter is practically negligible.
However, high energy photons penetrate more easily deeper layers of the
clouds so the radiation is thermalized more effectively and the UV part of
the reflected spectrum as well as the dark sides of clouds  are slightly 
brighter. 

More important role for the shape of the reflected spectrum is played by the
slope of the incident radiation.
In Fig. 6 we show two examples of model A (reflected spectrum) 
calculated for the energy index $\alpha$ equal to 0.9 and 1.1. Steeper
spectrum of the incident radiation results in steeper (i.e. softer) 
spectrum both in the optical band and in the soft X-rays.

\begin{figure}
\epsfxsize = 80 mm \epsfbox[50 180 530 660] {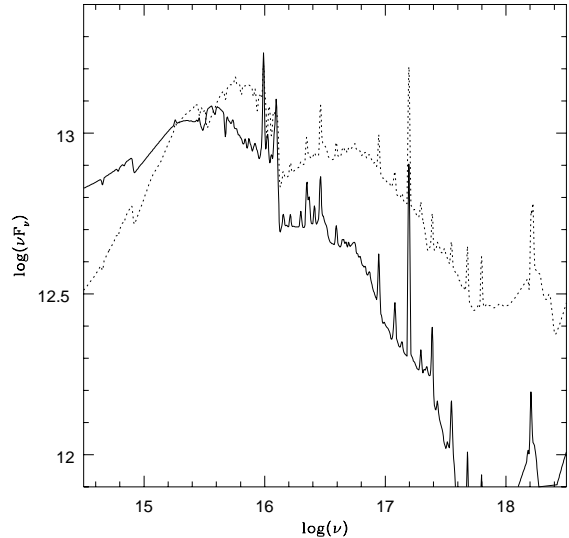}
\caption
{Two examples of the optical/X-ray spectrum coming from the bright sides of 
clouds calculated for model (A) assuming the slope $\alpha = 0.9 $ (dotted 
line) and $1.1$ (continuous line). 
Other parameters: ionization parameter $\xi = 1000$, extension of the incident
spectrum from 0.1 eV to 280 keV, cloud density $10^{12}$
cm$^{-3}$ and the column density $10^{26}$ cm$^{-2}$.} 
\end{figure}

\subsubsection{The nature of the primary emission}

In order to show the dependence of the OTCM model on the assumption about the
primary emission we present examples of solutions in which identical
distributions of clouds are exposed to primary emission with the same 
ionization parameter but different spectral properties. The shapes of the
incident spectra were shown in Fig. 3. 

In the IR/optical band model (A) is different from the other
two. If the covering factor of the source is not close to 1 the primary 
emission directly contributes to this spectral band and may strongly modify 
the optical slope
of the resulting spectrum (see also Section 3.2) making it softer. The 
visibility of the primary emission in that energy band also
affects the variability since in this case a fraction of the optical emission
is not expected to be delayed with respect to the X-ray band.

In the EUV/X-ray band model (C) is significantly different from the 
other two models. Power law models are featureless while the Comptonization of the
radiation emitted by the clouds preserves traces of atomic features if the
optical depth of the hot medium is lower than $\sim 0.5$. Therefore in 
synchrotron
models all line emission (in particular, Fe $K_{\alpha}$ line) has to be 
delayed with respect to the power law while in the last case the 
clear division of the X-ray emission into primary and reflected component is
difficult and a fraction of the line flux does not have to be delayed
with respect to power law.

\begin{figure}
\epsfxsize = 80 mm \epsfbox[50 180 530 660]{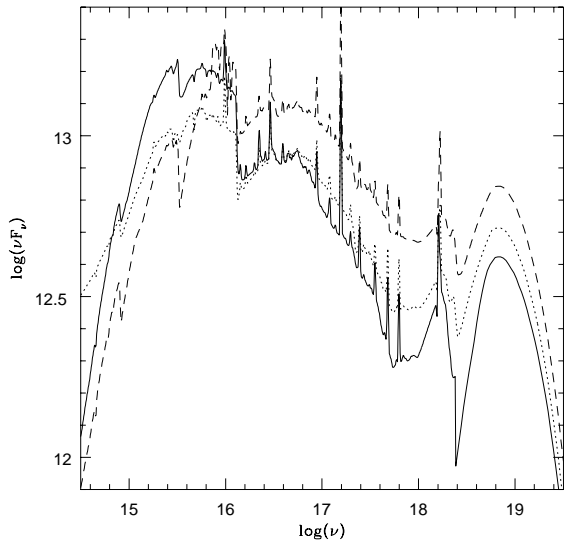}
\caption
{Three examples of the IR-Xray spectrum reflected from OTCM clouds 
calculated under 
assumptions of three primary emission models described in Section 2.2:
model (A) - dotted line, model (B) - dashed line, model (C) - continuous line.
In all
cases the ionization parameter of the clouds is equal to 1000, cloud density
is $10^{12}$ cm$^{-3}$ and the column density is  $10^{26}$ cm$^{-2}$.} 
\end{figure}

The extension of the primary emission into gamma band simply reflects the
adopted assumptions.

The spectra reflected by the clouds are shown in Fig.7. The emission lines
are included; they are plotted adopting a spectral resolution 
$\nu /\delta \nu$
equal to 30. 
The properties of the three spectra
are influenced by the choice of the primary emission mechanism.

The model (A) presented here differs with respect to the model given in Fig.
9a of Paper I mostly because of the change in the shape of the incident
radiation (slope of 0.9 instead of 1.0,  and the high energy cutoff (280 keV
instead of 100 keV). The new spectrum is generally harder because the incident
radiation is harder and elastically scattered photons constitute a significant
fraction of the reflected spectrum. The improvement
of the 
physical input into the code resulted in slightly stronger emission lines
and a more shallow Lyman edge.

The optical/UV slope of the reflected component is significantly flatter 
in the 
case of model (A) than in the other two. However, the traces of
a Balmer edge and of a significant Lyman edge can be seen in all these spectra.

The soft X-ray part of the reflected spectrum looks similar in all three 
models as it is mostly
determined by the adopted value of the ionization parameter. There is 
considerable
emission below 2 keV with an approximate photon index 2.5 
between 0.2 and 2 keV. 
The strong Fe $K_{\alpha}$ line is also characteristic for all three
spectra, but the  $K_{\alpha}$ edge is the weakest for model (A) and
the most profound for model (C).  

The broad band spectral index $\alpha_{ox}$, measured customary 
between  2500 \AA
~and 2 keV, for the reflected component itself is equal to 1.09 for model (A),
0.94 for model (B) and 1.14 for model (C). 

The high frequency part of the spectrum is again similar in all three cases
which is due to the fact that the shape of the reflection component at high
frequencies is simply determined by the Klein-Nishina cross-section for
scattering.

\begin{figure}
\epsfxsize = 80 mm \epsfbox[50 180 530 660]{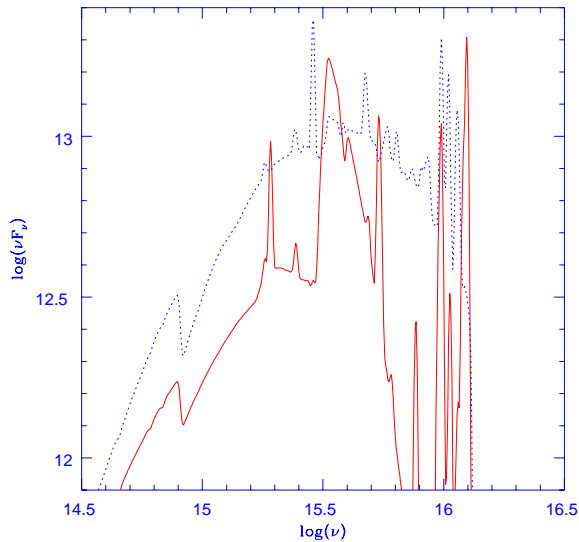}
\caption
{The emission of the dark sides of OTCM clouds 
calculated under 
assumptions of model (C) described in Section 2.2.1 (continuous
line). Dotted line shows a case (A) spectrum but with the temperature of the
dark side of the cloud equal to $5 \times 10^4$ K, higher than results from 
pure radiative heating.
The ionization parameter of the clouds is 
equal to 1000, cloud density
is $10^{12}$  cm$^{-3}$ and the column density is $10^{26}$ cm$^{-2}$.} 
\end{figure}

An example of radiation spectra emitted by the dark sides of the clouds 
is shown in Fig. 8. The figure shows case (C) but the other two spectra
are also characterized by an extremely large Lyman discontinuity in emission
unless an extra heating of the dark sides of the clouds is allowed (dotted
line). 

For these spectra  the $\alpha_{ox}$  
index is high in all cases since the contribution of the dark side of the
clouds to the spectrum at 2 keV is negligible. Therefore any contribution 
of the dark sides of the clouds to the resulting spectra would give steeper 
$\alpha_{ox}$
than pure reflection spectra. Any contribution from primary emission would work
in the opposite direction, and additionally it would modify the soft X-ray
slope.

\subsection{Optical/UV slope}

\begin{figure}
\epsfxsize = 80 mm \epsfbox[50 180 530 660]{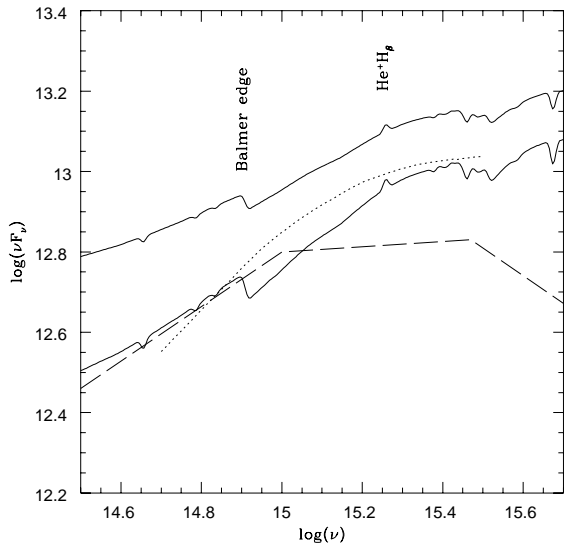}
\caption
{Optical/UV  spectrum of OTCM clouds 
calculated under 
assumptions of emission model (A) described in Section 2.2.1. The lower curve
shows the reflected spectrum and the upper curve shows a case of 
an equal contribution
from the reflected and primary emission. The ionization parameter 
of the clouds is 
equal  to 1000, cloud density
is $10^{12}$  cm$^{-3}$ and the column density is $10^{26}$ cm$^{-2}$
The dashed line shows the mean spectrum of radio quiet quasars from Laor et 
al. (1997) and the dotted line shows the mean quasar spectrum of Francis et al.
(1991).} 
\end{figure}

We show an expanded fragment of the spectrum calculated from model (A) in Fig.
9. We consider two extreme cases interesting from the observational point of 
view. The lower curve shows a pure reflection component, i.e. corresponds to the case when the primary emission is not seen directly because it comes from a
compact source and is shadowed by one of the clouds. The upper curve was 
obtained assuming 
that half of the primary source emission reaches the observer. 

In both cases we see the characteristic atomic features: noticeable Balmer 
edge, Lyman edge and weak $He^+H_{\beta}$ line. Equally profound spectral features 
are seen in models (B) and (C) since the spectral shape of the incident 
radiation  has little effect. The Comptonization within the clouds 
is not expected to remove these features for the clouds of the adopted 
properties and $\xi = 1000$ (Paper I).

The clear difference between model (A) and models (B) and (C) is in the
slope of this part of the spectrum. We compare the models with the mean 
spectrum for radio quiet quasars derived recently by Laor et al. (1997). Our
case (A) nicely coincides with the data in the optical part (if primary and 
the dark sides of the clouds are  
not seen) but is far too bright in the UV which means that the temperature of 
the clouds is too high. In the case of model (B) and (C) the reflection 
spectra 
are definitely too hard in the optical band and any contribution from the
primary emission does not change that. 

The comparison with the mean
quasar spectrum obtained by Francis et al. (1991) is more favorable for
model (A) and in that case a minor contribution from the primary emission
would even be allowed. However, models (B) and (C) are still too hard to match
observations for the ionization parameter $\xi = 1000$.  

Any contribution from the emitted spectra would give still harder spectra,
in contradiction with the data.

The choice of lower value of the ionization parameter $\xi$ results in a lower
value for the cloud temperature which in principle helps to reconcile the 
models (B) and (C) with the data. We computed the spectra assuming $\xi=300$.
However, in that case the spectral features (Balmer edge in absorption 
and Lyman edge in emission) are strong, giving a change in the continuum by
50 \% and 100 \%, respectively, in the case of a reflection component and
even more (100 \% and 500 \% respectively) in the case of emission from
the dark sides of the clouds.

\subsection{Soft X-ray emission}

\begin{figure}
\epsfxsize = 80 mm \epsfbox[50 180 530 660]{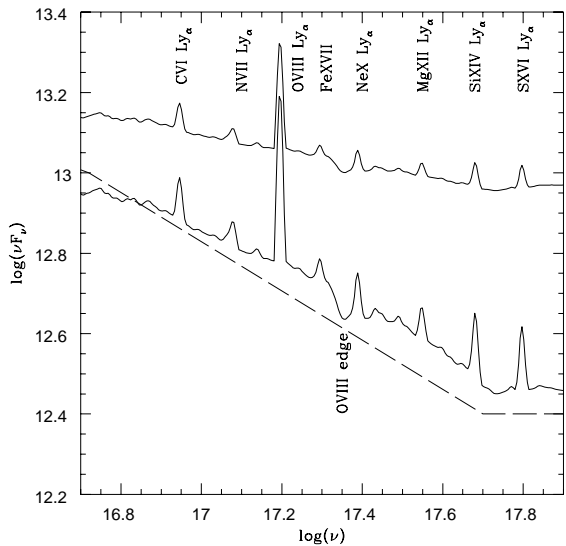}
\caption
{Soft X-ray spectrum of OTCM clouds 
calculated under 
assumptions of emission model (A) described in Section 2.2.1. The lower curve
shows the reflected spectrum and the upper curve shows the observed spectrum
for the weight from primary 0.5. The ionization parameter of the clouds is 
equal to 1000, cloud density
is $10^{12}$ cm$^{-3}$ and the column density is $10^{26}$ cm$^{-2}$.
The dashed line shows the mean spectrum of radio quiet quasars from Laor et 
al. (1997).} 
\end{figure}

Since the spectrum emitted by the dark sides does not contribute to the 
soft X-rays we can restrict
our study of that band to the reflected component and the contribution from
the primary emission.

We show an expanded version of model (A) in Fig. 10, in two versions: 
pure reflection spectrum and reflection plus half of the primary emission. We
assumed $\xi=1000$.

The overall shape of the spectrum looks like a power law. When pure reflection
is considered the slope of this power law is similar to the mean spectrum of
radio quiet quasars (Laor et al. 1997). If a significant contribution from the primary is
allowed the slope is too flat. Models (B) and (C) are generally 
similar in this spectral
band so they are also an adequate description of the data.

Precise values of the slope between 0.2 keV and 2 keV for the reflected 
component in the three models are equal to 1.48, 1.35 and 1.57, 
respectively.
The slope in the Laor et al. (1997) sample is 1.6 so model (C) is the
best representation of the data. Even in the last model there is no place for
a significant contribution from the direct primary emission since the slope
would be too flat. 

We see a number of quite strong emission lines, with OVIII at 0.653 keV being
the most prominent. Complex spectral features are observed in a number of 
sources both in absorption and in emission, and they are usually 
explained by warm 
absorbers, i.e. optically thin clouds with a column density $\sim 10^{23}$
and located at a distance of several light weeks (e.g. Otani et al. 1996). 
The same 
features are seen in models (B) and (C). Neither the mean quasar spectrum of 
Laor et al. (1997) nor observations of sigle quasars (e.g. Leach et al. 
 1995 for 3C 273) indicate the presence of such features.

A higher value of the ionization parameter would give too flat slopes for all
three models.

Lower values of the ionization parameter  give systematically
steeper slopes. Model (A) calculated for $\xi = 300$ requires a
contribution of the primary emission with a weight of 
0.2  in order to reproduce
the observed slope. For model (C) the allowed contribution of the primary is
always slightly higher for a given value of the ionization parameter, 
and for $\xi = 500$ the required weight of the primary is 
$\sim 0.15$. The line emission is much stronger in that case.

In any case, the OTCM model for quasars explains the soft X-ray emission as a
reflection from a partially ionized gas. The same mechanism was suggested
to produce weak soft X-ray excesses in a number of Seyfert 1 galaxies (Czerny
\& \. Zycki 1994). However, in their case there were some problems to
achieve the required ionization state of the gas within the frame of the 
adopted
model. No such problems are encountered here.

\begin{figure}
\epsfxsize = 80 mm \epsfbox[50 180 530 660]{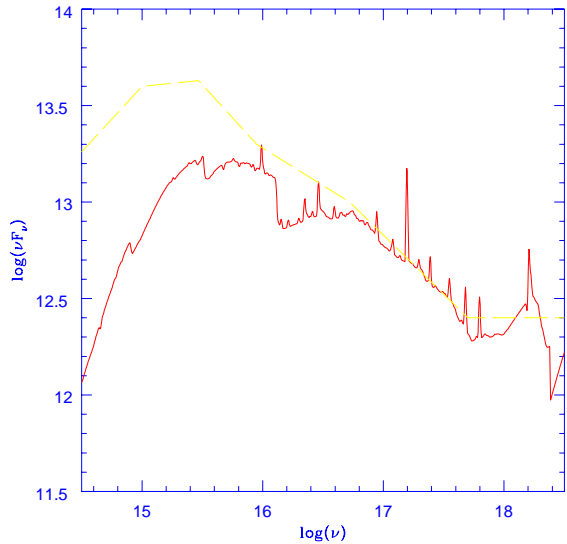}
\caption
{Optical/UV/soft X-ray  spectrum of OTCM clouds for  pure reflection 
calculated under 
assumptions of emission model (C) described in Section 2.2.1 (continuous
line).  The ionization parameter of the clouds is 
equal to 1000, cloud density
is $10^{12}$ cm$^{-3}$ and the column density is $10^{26}$ cm$^{-2}$.
The dashed line shows the mean spectrum of radio quiet quasars from Laor et 
al. (1997).} 
\end{figure}


\begin{figure}
\epsfxsize = 80 mm \epsfbox[50 180 530 660]{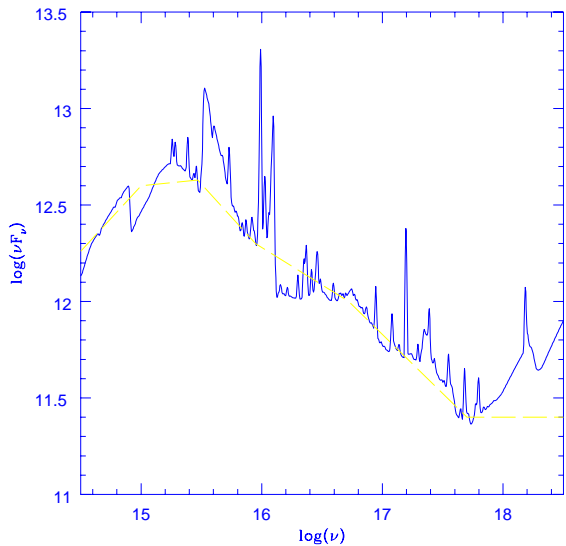}
\caption
{Optical/UV/soft X-ray  spectrum of OTCM clouds 
calculated under 
assumptions of emission model (A) described in Section 2.2.1 (continuous
line) and assuming the weight of the reflected radiation equal to 1.0, of the
dark sides emission 1.2, and of the primary  0.15. 
The ionization parameter of the clouds is 
equal to 300, cloud density
is $10^{12}$  cm$^{-3}$ and the column density is $10^{26}$ cm$^{-2}$.
The dashed line shows the mean spectrum of radio quiet quasars from Laor et 
al. (1997).} 
\end{figure}

\begin{figure}
\epsfxsize = 80 mm \epsfbox[50 180 530 660]{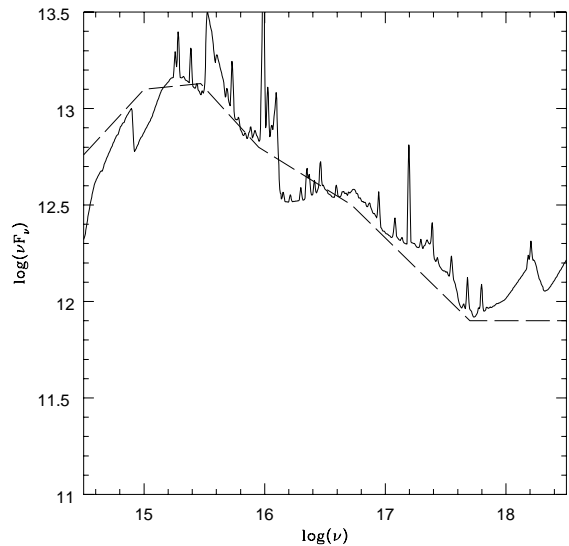}
\caption
{Optical/UV/soft X-ray  spectrum of OTCM clouds 
calculated under 
assumptions of emission model (C) described in Section 2.2.1 (continuous
line) and assuming the weight of the reflected radiation equal to 1.0, of the
dark sides emission 5.0, and of the primary  0.15. 
The ionization parameter of the clouds is 
equal to 500, cloud density
is $10^{12}$  cm$^{-3}$ and the column density is $10^{26}$ cm$^{-2}$.
The dashed line shows the mean spectrum of radio quiet quasars from Laor et 
al. (1997).} 
\end{figure}


\subsection{Matching soft X-ray emission with UV}

So far we discussed the comparison of the models with the mean quasar 
spectrum in the optical/UV band and in the 
soft X-ray band separately, i.e. with an 
arbitrary normalization in both bands. The comparison of the data with the
model in the entire optical/UV/soft X-ray band is shown in Fig. 11 (model C,
pure reflection). If the normalization is adjusted to soft X-rays we 
notice that the observed flattening towards high X-rays (at $\sim 2$ keV) is
reproduced by the models. 

The $\alpha_{ox}$ index of the mean quasar spectrum of Laor et al. (1997) is
equal to 1.46 while in the reflected spectra shown in Fig. 7 it is equal
to 1.09, 0.94 and 1.14, respectively. However, a contribution of the 
emission of the dark sides of the clouds helps to fill the gap.

Since the temperature of the dark side of a cloud depends significantly
on the value of the ionization parameter $\xi$ we can adjust its value
to model the UV/EUV spectrum (apart from the spectral features present in 
the model and absent in the data).

In the case of model (A) the best value of the ionization parameter is
$\sim 300$ and an example of the resulting spectrum is shown in Fig. 12. 
Even the slight bend below $\sim $0.2 keV is reproduced by the
models. 
The contribution of the
dark sides of the clouds required to provide the flux in the optical band is
rather moderate and the entire spectrum is mostly dominated by reflection. 
In the case of model (C) the best value of the ionization parameter is slightly
higher, $\sim 500$ (see Fig. 13). However, there is a large deficit of emission
in the optical band which shows that a number of additional, cooler clouds 
should be included in the model, i.e. a single cloud  population located at
a given distance (i.e. parametrized by a single value of ionization parameter)
is not  an adequate description of the data in the case model (C).

The spectral features in the optical/UV band are quite large
in that case. The kinematic broadening of these features due to the motion
of the clouds may slightly reduce and broaden those features (see Fig. 15 
for this
effect in the soft X-ray band) 
but will not really 
remove them and we intend to devote a special paper to that problem since 
it is one of the major issues in all realistic models, including 
accretion disks.

\subsection{High energy tail and the level of primary emission}

The high energy cut-off of the spectrum is not well constrained observationally
in the case of radio quiet quasars. Therefore, it is difficult to formulate
any preferences for any of the discussed models on the basis of this 
spectral band.

\begin{figure}
\epsfxsize = 80 mm \epsfbox[50 180 530 660]{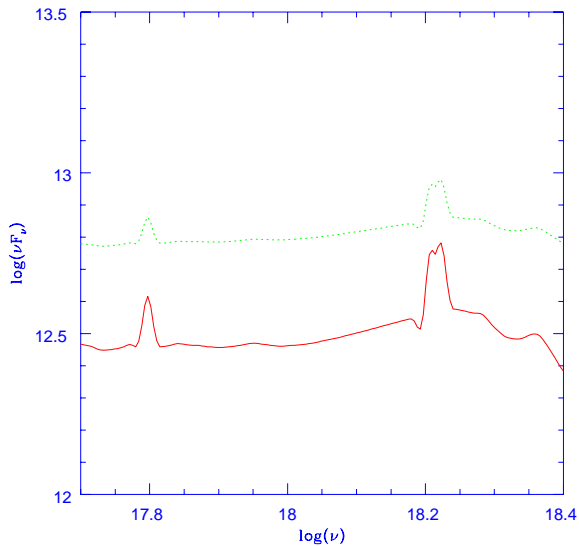}
\caption
{The spectrum of OTCM clouds 
calculated under 
assumptions of emission model (A) described in Section 2.2.1 (continuous
line). Dotted line shows a case (A) spectrum with the contribution from 
the primary with the weight 0.5.
The ionization parameter of the clouds is 
equal to 1000, cloud density
is $10^{12}$  cm$^{-3}$ and the column density is $10^{26}$ cm$^{-2}$.} 
\end{figure}

The intensity of the iron $K_{\alpha}$ line given by the model is clearly 
large, in contradiction with observations of quasars, since a significant
contribution from the primary emission is not allowed. For $\xi = 1000 $ 
the EW is 
$\sim 1200$ eV if the
direct primary emission is not seen (lower curve in Fig. 14) but it is
reduced to $\sim 240 $ eV if the contribution with a weight 0.5 would
be 
allowed (upper curve in Fig. 14). For lower $\xi$ it is still larger, 
and equals $\sim 2.3 $ keV and
$\sim 460$  eV, respectively. This is not surprising if the reflected
spectrum dominates since the line intensity expected in such case is large
(e.g. \. Zycki \& Czerny 1994). 

Observations of moderately bright quasars (Nandra et al. 1997b) give the 
equivalent width of the iron $K_{\alpha}$ line not higher than 300 eV, with 
a clear trend for a decrease of this value with an increase of a quasar
luminosity.  

\section{Discussion} 

\subsection{Nature of primary emission}

We used our OTCM model from Paper I in order to analyze the requirements 
which result from the comparison of the three variants of the model to Laor 
et al. (1997) mean spectra of radio quiet quasars. The advantage of 
concentrating on quasars was in a reliable description of their broad band 
spectra from optical band to hard X-rays. 

Model (A) reproduces well the entire broad band spectrum of quasars, 
including the optical slope of the data.
On the other hand, present computations carried for that model allow only a
minor contribution from the primary emission which results in a too high 
equivalent width of the iron $K_{\alpha}$ line and a need of some special
geometry for cloud distribution in order to explain this level of primary
emission (see Section 4.3). 

Model (B) is the least attractive since it displays the same problems as
model (A) with respect to the level of primary emission and it does not
explain the optical slope of quasars.

Model (C) is also too steep (i.e. too hard) in the optical band in comparison
with the mean quasar spectrum although it models well the UV/X-ray 
band. This may mean that model (C) well represents well the innermost part of
the accretion flow but it has to be supplemented by the presence of more distant
clouds and/or an outer accretion disk. However, this model also display the
problem of too high equivalent width of the iron $K_{\alpha}$ line since it
again allows only a marginal contribution of the direct primary emission.

\subsection{Ionization level}

The modelling of the overall shape of the spectrum was very sensitive to
the choice of the ionization parameter $\xi$ since this parameter
determines the cloud temperature and the emission from their dark sides
which dominate the optical/UV part of the spectrum.

Our analysis based on the overall spectral shape clearly favored an $\xi$ of
order of 300-500 to fit the mean quasar spectrum of Laor et al. (1997). The
value $\xi = 1000$ gave a cloud temperature too high to fill the gap in the
optical band between
the model predictions and the observed spectra if the soft X-ray band was
modelled correctly. It also left no room for any presence of the primary
emission since the soft X-ray slope of the reflected component was just
marginally steep enough for model (C) to match the data.
 
We can compare this result with more detailed spectroscopic constraints.

The most direct estimate of the ionization level of the cool gas in quasars
comes from the analysis of the position of the $Fe K_{\alpha}$ line. ASCA
results show that for moderately bright quasars the line is at 
6.4 - 6.5 keV, while for bright quasars it is at 6.57 keV 
(Nandra et al. 1997b). 
This suggests that the ionization parameter for bright quasars is of the order of
a few hundreds (e.g. \. Zycki \& Czerny 1994). In fact, the flux averaged 
position of the iron line in our computations is 6.82 keV for $\xi=1000$,
model (A), 6.43 keV for $\xi=300$, and 6.61 keV for $\xi=500$, model (C). 
Therefore, also from this point of view, 
all intermediate value of $\xi$ better corresponds to the observations.

\subsection{Geometry of clouds distribution and the level of primary emission}

In the case of purely random distribution of clouds with a covering factor
$\Omega/ 4 \pi$ and averaged observing angle the relative contribution of the
primary emission should also be determined by the covering factor, thus
giving a formula appropriate for modelling a mean spectrum:

\begin{equation}
F_{obs} = (1 - {\Omega \over 4 \pi})F_i +{\Omega \over 4 \pi}F_{dark} +
{\Omega \over 4 \pi}(1 - {\Omega \over 4 \pi})f_{ampl}F_r 
\end{equation}
where $f_{ampl}$ is the amplification factor taking into account the multiple
scattering 
\begin{equation}
f_{ampl}= ({1 - {A\Omega \over 4 \pi}})^{-1}
\end{equation}
where A is the effective albedo approximately equal 0.85 (see Paper I). 
This formula gives the maximum
efficiency since it assumes that the source of primary emission does not
intercept repeatedly scattered radiation. 
Other quantities have the
same meaning as in Paper I, i.e. $F_i$ is the primary (incident) radiation,
$F_{dark}$ is the emission of the dark sides of the clouds and $F_r$ is the
radiation reflected/reemitted by the bright sides of the clouds. 
In the case of model(C) the amplification factor is decreased by the presence
of hot plasma intercepting a fraction of the photons and the appropriate 
formula is given by
\begin{equation}
f_{ampl}= (1 - A(1 - ({R_X \over R_{UV}})^2)){\Omega \over 4 \pi}\bigr)^{-1}
\end{equation}

Such an approach reduces the number of the original free parameters of the
model as all weights are now expressed by the covering factor 
$\Omega /4\pi$ which allows us to test the consistency of the derived model
parameters with the random distribution of clouds.

In the case of model (A) and $\xi = 300$ the Laor et al. (1997) data were 
reproduced by the
model with the relative weights of the components in Eq. (2) equal
to 0.15, 1.2 and 1.0. If we rely on the relative weight of 
the second and third term, this result can be translated into a
covering factor $\Omega /4 \pi = 0.57$. The role of the amplification is 
taken into account ($f_{ampl}=1.9$). Such a cloud distribution, 
while reproducing precisely the 
normalization of the last two components, predicts a significantly larger 
value of the
primary contribution (0.90 instead of 0.15). Lower amplification still 
widens this gap. This means that either the cloud distribution is not
random or our theoretical spectra are too hard (i.e. flat) in the soft X-ray 
band. Since the equivalent width of the Fe $K_{\alpha}$ line produced by the
model is also too large if the primary level is as low as 0.15 it strongly
suggests that the geometry is correct but the reflection spectra should be
improved.

The size of the primary 
source could in
principle be estimated from the statistical properties of the observational
sample of objects. We can only infer that the source is probably not too
compact in comparison with the clouds distance from the center since a point 
like 
source would give two distinct classes of objects: a fraction of objects 
dominated
by primary emission and the rest of the sources with the primary completely hidden 
from an observer, which seems not to be the case. 
 
In the case of model (C) a slightly higher value of $\xi$ seems to be favored,
of order of 500.
The relative weights of the components  are equal to 0.15, 5.0 and 1.0. Therefore,
dark sides of the clouds are contributing more to the total spectrum.
However, now the amplification is weaker by a factor $1- (R_X/R_{UV})^2$ due 
to the 
size of the hot cloud. Combining the requirements for the relative contribution
of the reflected component, dark side component and equation (1) we obtain
$\Omega /4 \pi = 0.88$  and $R_X/R_{UV}=0.66$. Such
a model predicts the contribution from the primary (hot plasma
in that case), of order of 0.65, a factor of 4 higher than allowed by the 
data.

It is interesting to note that the condition for the size of the hot plasma
cloud (eq. 1) leads to a reasonable value of the $R_X/R_{UV}$ ratio for the
value of the Compton parameter $y$ which resulted from the choice of the
plasma parameters giving the hard X-ray slope of index $\sim 0.9$.

\begin{figure}
\epsfxsize = 80 mm \epsfbox[50 180 530 660]{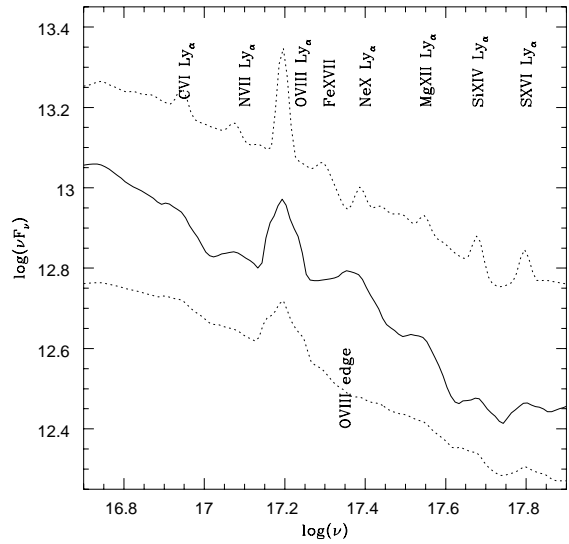}
\caption
{The importance of the kinematics for the spectral features in the 
soft X-ray spectrum of OTCM clouds. The upper dotted line shows the 
spectrum from
Fig. 10 (model A) but  coming from clouds located
at randomly oriented orbits of the radius $50 R_{Schw}$, the lower dotted curve
shows the same spectrum but from clouds circulating at $7 R_{Schw}$ and the 
continuous line shows model (C) from Fig. 13 but coming from clouds at $12
R_{Schw}$. The spectra were systematically shifted for convenience.} 
\end{figure}

The model specifies the $R_X/R_{UV}$ ratio, but not $R_{UV}$ itself. However,
if we assume that the mean quasar spectrum actually corresponds to a mean
value of the accretion rate $2.8 M_{\odot}$yr$^{-1}$ (i.e. total 
luminosity $10^{46}$ erg s$^{-1}$) and a mean value of the
black hole $1.4 \times 10^9 M_{\odot}$ (after Zheng et al. 1997) 
the adopted value of the 
cloud density
and the obtained value of the ionization parameter $\xi = 500$ suggest 
immediately
that the  representative value of $R_{UV}$ is of order of  $12 R_{Schw}$. We
cannot fully rely on the value of the accretion rate and the mass of the black
hole since they depend on the choice of a specific model.
However, any reasonable assumption about the luminosity to the Eddington
luminosity ratio (of order of 0.1 in this case) would give a similar order of 
magnitude. Therefore, the 
kinematic effects are important. In Fig. 15 we show how the soft X-ray
emission features are affected by the motion of the clouds. We see that the
result is quite sensitive to the actual distances of the clouds from the
center. Larger distances, of about $50 R_{Schw}$ leave most of the spectral
features almost unaffected while considerably smaller values, $7 R_{Schw}$,
smear out all the features apart from a strong OVIII line which is very broad.
Therefore, if soft X-ray emission is really produced as a reflection spectrum
the presence and the strength of the spectral features should help to test
the model. Unfortunately, the available data for quasars are not yet of the
appropriate quality and there is an additional problem of confusion with the
spectral features due to warm absorber. 

\subsection{Future prospects}

The comparison between the theoretical models and the data shown in Sect.
4 did not include spectral fitting in terms of determination of the
$\chi^2$ statistics since the cloud model is not ready yet 
for that kind of quantitative analysis. Clearly, very careful, more advanced
computations of both the reflected spectrum and the emission of the dark side
of a cloud are necessary in order to decide whether the soft X-ray part of
the quasar spectrum can be identified with the radiation reflected by the
irradiated sides of the clouds, with some contribution from the primary 
radiation. 

Actually, preliminary computations taking into account multiple scattering
(see Paper I) indicate that the amplification is wavelength-dependent as the
spectra are steeper in soft X-rays due to this effect. More careful 
computations based on radiative transfer of soft photons and Monte Carlo 
computations of X-ray photons show that the slope of the spectrum for 
$\xi = 300$ has the slope as steep as 2.2 for the reflection component in 
the 0.2
- 2 keV band due to the multiple reflection (Abrassart et al., 
in preparation). 
Such a steep reflection 
spectrum would allow a primary contribution as high as 0.5 which is 
only lower by a factor two than expected from a random distribution of the 
clouds.

The geometrical arrangement of the origin of the primary radiation
does not seem equally important as the physical input of the radiative
transfer code since the geometrical parameters will adjust themselves to the  
shapes of the basic spectral components.

Considerable help may be provided by the variability constraints. In the OTCM
there are three kinds of variations expected. 

The first one is connected with
variations of the primary source, or the comptonizing hot medium. Clouds 
respond to those variations both in optical/UV and in soft X-rays with an 
average delay of order of the travel time
through the region they occupy, $R_{UV} \sim 10 R_{Schw}$, of order of some
ten thousand of seconds for a $10^8 M_{\odot}$ black hole. 

The second one is
connected with a single cloud motion, i.e. the visibility of the primary.
These variations are mostly limited to the primary variations and some 
changes in optical/UV or soft X-rays resulting from the temporary change in
geometry are not expected to be delayed in any systematic way. 
The time-scale of these variations 
should be of order of the cloud period diminished by the factor describing
the relative size of the X-ray source, $R_X/R_{UV}$, thus not considerably 
longer than the previous time-scale. Variations in quasars on those time-scales
are unmeasurably weak in the optical band (below 1.2 \% in time-scales of hours in
3C 273, von Montigny et al. 1997).  

Finally, any systematic changes in the
accretion rate in the innermost part of the flow which would both include a
change of the level of primary emission as well as of the covering factor should
happen on considerably longer time-scales. However, at the present stage 
the OTCM
does not give quantitative predictions of the relative changes of these two
factors, although we generally expect that the covering factor should grow
with accretion rate, i.e. with bolometric luminosity, thus resulting in
harder UV spectra and larger $\alpha_{ox}$ for larger luminosity. 

The variability in hard X-ray band for radio quiet quasars also seems to be 
the best and most 
direct probe of the nature of the primary source. High quality
data for galactic sources allow us to compute the time delays as functions of the
Fourier phase as well as the coherence function (e.g. Cui et al. 1997) 
which strongly support the
Comptonization mechanism for the production of 'primary emission' and 
provide a potentially
powerful method to constrain the distribution of the hot gas (e.g. Hua, 
Kazanas \& Titarchuk 1997). 
The time-scales
involved are in the range of $10^{-3}$s to 1s for Cyg X-1 so simple minded scaling
may suggest time-scales from days to years.

\section{Conclusions}

Our results suggest that the quasar spectra can be explained within
the frame of the optically thick cloud model (OTCM), i.e. as originating in a 
distribution of irradiated 
clouds optically thick for electron scattering. 
One of the important points is that the spectrum in the  soft X-ray 
band is attributed mostly 
to the reflection from the illuminated partially ionized
sides of the clouds. We are not able to definitively distinguish at present 
if the
primary emission comes from the unspecified central source in the form of
synchrotron emission or if it originates in the central hot plasma cloud by
upscattering of soft photons coming from the clouds. However, we favor the second
possibility as the presented results already lead to more self-consistent
description for that case.

An alternative picture based
on the accretion disk and a corona requires very hot plasma to form the hard
X-ray emission and moderately hot but optically thicker plasma to explain
the soft X-ray quasar spectra. These two pictures differ with respect to
the presence of spectral features since the reflection is accompanied by emission
lines and absorption edges which may be used for the purpose of diagnostics. 
These features are expected to be smeared to some extent by
the cloud motion which may help to distinguish them from features arising
in a distant warm absorber.

Within the frame of OTCM, the overall shape of the spectrum is very sensitive
to the value of the ionization parameter $\xi$. Interestingly, the value
favored by the comparison of the model to the mean quasar spectrum of Laor
et al. (1997) is $\sim 500$, 
in agreement with the
energy of the Fe $K_{\alpha}$ line in moderately bright quasars (Nandra et al.
1997b). Present results allow too low a contribution (by a factor of a few) 
from the primary emission
to account for the observed equivalent width of $K_{\alpha}$ but we expect
that more advanced computations may remove this problem (Abrassart et al.,
in preparation). 

The problem which remains is the prediction of the  strong spectral 
features in the
optical/UV band. Observations seem to indicate that no such features are
observed in quasar spectra while both OTCM and advanced accretion disk
models predict such atomic features although kinematic effects connected with 
the cloud motion decrease them to some extent. We will address this problem in
the future.

\begin{acknowledgements}
We are grateful to Suzy Collin for many helpful discussions and 
detailed comments on the manuscript. We thank Piotr \. Zycki for his help
with the high energy reflection code. 
This work was partially supported  
by grant 2P03D00410 of the Polish State Committee for 
Scientific Research and by Jumelage/CNRS No 16 ``Astronomie France/Pologne''. 
\end{acknowledgements}

\end{document}